\documentclass[a4paper, oneside, 11pt]{report}
\usepackage[utf8]{inputenc}
\usepackage[T1]{fontenc}
\usepackage{graphicx}
\usepackage{longtable}
\usepackage{colortbl}

\usepackage{xcolor}
\usepackage[linesnumbered,ruled,vlined]{algorithm2e}

\SetCommentSty{mycommfont}

\SetKwInput{KwInput}{Input}                % Set the Input
\SetKwInput{KwOutput}{Output}              % set the Output

\usepackage{pdfpages}
\usepackage[nottoc,notlot,notlof]{tocbibind}
\usepackage{titlesec}
\titleformat{\chapter}[display]{\bfseries\centering}{\LARGE Chapter \thechapter}{0em}{\LARGE}

\titlespacing*{\chapter}{0pt}{-20pt}{20pt}

\usepackage[hidelinks]{hyperref}
\usepackage{float}

\usepackage{caption}

\usepackage[left=1.5cm, right=2cm, top=2.5cm, bottom=2.5cm, bindingoffset=0.5cm, head=10pt]{geometry} 
\usepackage{setspace}
\usepackage{fancyhdr}
\usepackage{float}
\usepackage{enumitem}
\usepackage{wrapfig}
\usepackage{amsmath}
\numberwithin{figure}{section}

\usepackage{titlesec}
\titleformat{\paragraph}
{\normalfont\normalsize\bfseries}{\theparagraph}{1em}{}
\titlespacing*{\paragraph}
{0pt}{3.25ex plus 1ex minus .2ex}{1.5ex plus .2ex}

\usepackage{caption}
\usepackage{subcaption}
\usepackage{nomencl}
\makenomenclature

\title{Locality-based Graph Reordering for Processing Speed-Ups}

\begin{document}

%%%%%%%%%%%%%%%%%%%%%%%%%%%%%%%%%%%%%%%%%%%%%%%%%%%%%%%%%%%%%
%TITLE PAGE (Pre-defined, just change parameters above)
%%%%%%%%%%%%%%%%%%%%%%%%%%%%%%%%%%%%%%%%%%%%%%%%%%%%%%%%%%%%%
%%%%%%%%%%%%%%%%%%%%%%%%%%%%%%%%%%%%%%%%%%%%%%%%%%%%%%%%%%%%%
%TITLE PAGE
%%%%%%%%%%%%%%%%%%%%%%%%%%%%%%%%%%%%%%%%%%%%%%%%%%%%%%%%%%%%%
\makeatletter
\begin{titlepage}
    \begin{center}
        \vspace*{1cm}

        \Large{
        \textbf{Locality-based Graph Reordering for Processing Speed-Ups and Its Dependence on Graph Diameter}}

        \vspace{1.5cm}
        
        %\thesistype{}
       \normalsize{\textbf{Bachelors' Thesis Project}\\
        Submitted in fulfillment of the requirements for the degree of \\
        \textbf{Bachelor of Technology}\\
        in \\
        Electrical Engineering}
        
        \vspace{1cm}

        \begin{figure}[htbp]
             \centering
             \includegraphics[width=.5\linewidth]{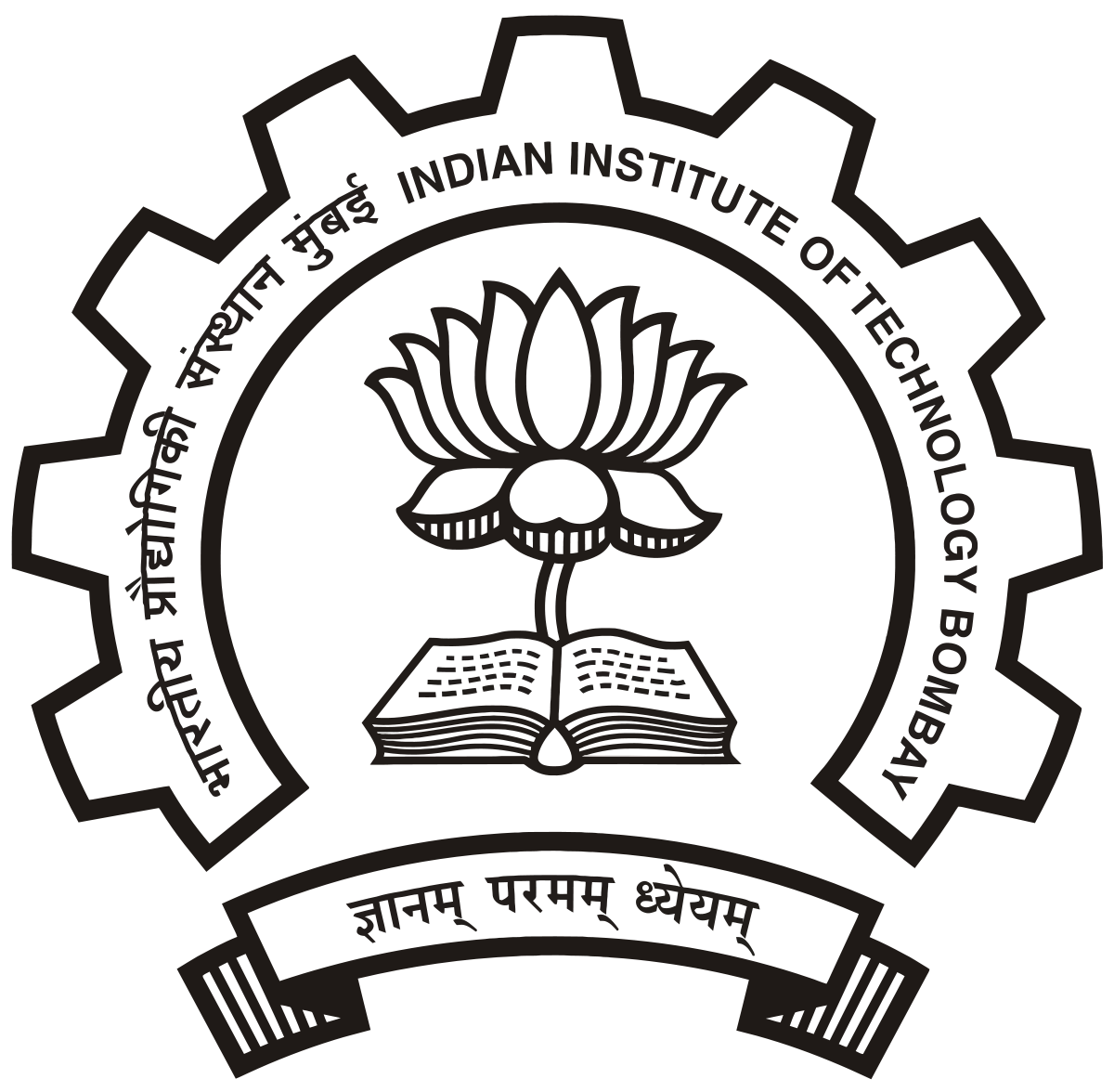}
        \end{figure}

        \vspace{1cm}

        \large
        \textbf{Author}: Vedant Satav (Roll No. 170070012)\\
        \large
        \textbf{Guide}: Prof. Virendra Singh

        \large
        
        \vspace{1cm}
        \large
        Department of Electrical Engineering\\
        
        Indian Institute of Technology Bombay\\

        \vspace{1cm}
        July 4th, 2021

    \end{center}
\end{titlepage}
\makeatother

%%%%%%%%%%%%%%%%%%%%%%%%%%%%%%%%%%%%%%%%%%%%%%%%%%%%%%%%%%%%%
%SOOA
%%%%%%%%%%%%%%%%%%%%%%%%%%%%%%%%%%%%%%%%%%%%%%%%%%%%%%%%%%%%%
%\input{Template/dissertation}
%\includepdf[pages={1}]{Template/Approval_Sheet_160070031.pdf}

%%%%%%%%%%%%%%%%%%%%%%%%%%%%%%%%%%%%%%%%%%%%%%%%%%%%%%%%%%%%%
%ABSTRACT
%%%%%%%%%%%%%%%%%%%%%%%%%%%%%%%%%%%%%%%%%%%%%%%%%%%%%%%%%%%%%
\clearpage
\thispagestyle{empty}

\printnomenclature
%%%%%%%%%%%%%%%%%%%%%%%%%%%%%%%%%%%%%%%%%%%%%%%%%%%%%%%%%%%%%
%TOC,TOF,TOT
%%%%%%%%%%%%%%%%%%%%%%%%%%%%%%%%%%%%%%%%%%%%%%%%%%%%%%%%%%%%%
\clearpage

\doublespacing

\section*{Abstract}
Graph analysis involves a high number of random memory access patterns. Earlier research has shown that the cache miss latency is responsible for more than half of the graph processing time, with the CPU execution having the smaller share. There has been significant study on decreasing the CPU computing time for example, by employing better cache prefetching and replacement policies. In this paper, we study the various methods that do so by attempting to decrease the CPU cache miss ratio.\\ \\
Graph Reordering attempts to exploit the power-law distribution of graphs- few sparsely-populated vertices in the graph have high number of connections- to keep the frequently accessed vertices together locally and hence decrease the cache misses. However, reordering the graph by keeping the hot vertices together may affect the spatial locality of the graph, and thus add to the total CPU compute time. Also, we also need to have a control over the total reordering time and its inverse relation with the final CPU execution time\\\\
In order to exploit this trade-off between reordering as per vertex hotness and spatial locality, we introduce the light-weight Community-based Reordering. We attempt to maintain the community-structure of the graph by storing the hot-members in the community locally together. The implementation also takes into consideration the impact of graph diameter on the execution time. We compare our implementation with other reordering implementations and find a significantly better result on five graph processing algorithms- BFS, CC, CCSV, PR and BC. Lorder achieved speed-up of upto $7\times$ and an average speed-up of $1.2\times$ as compared to other reordering algorithms\\\\
Index terms-- graph analysis, power-law distribution, ground-truth communities

%  \cite{ref_1_Intel_tb}

\clearpage
% \input{Template/acknowledgments}

% %\renewcommand{\nomname}{List of Abbreviation}
% \printnomenclature
% %%%%%%%%%%%%%%%%%%%%%%%%%%%%%%%%%%%%%%%%%%%%%%%%%%%%%%%%%%%%%
% %TOC,TOF,TOT
% %%%%%%%%%%%%%%%%%%%%%%%%%%%%%%%%%%%%%%%%%%%%%%%%%%%%%%%%%%%%%
% \clearpage
\pagenumbering{Roman}
\tableofcontents
\clearpage
%\listoffigures
\clearpage
%\listoftables
%\listofalgorithms
\clearpage

\pagenumbering{arabic}

%%%%%%%%%%%%%%%%%%%%%%%%%%%%%%%%%%%%%%%%%%%%%%%%%%%%%%%%%%%%%
%MAIN PART
%%%%%%%%%%%%%%%%%%%%%%%%%%%%%%%%%%%%%%%%%%%%%%%%%%%%%%%%%%%%%

% SEC1
\clearpage
\chapter{Introduction}
%\vspace{1.8em}
\label{sec:Intro}

Over the years, the applications of graph analysis have diversified, one among the many being the field of social network processing.\\
However, because of the way graphs are stored in the shared memory, their processing involves inherent irregular memory access patterns. The random accessing is further exacerbated by the varied processing style adopted by different algorithms used in graph processing. These memory indirections result in frequent cache-misses thereby leading to longer CPU cache latency issues and thus, limiting the efficiency of graph processing in shared memory systems.\\
One way to reduce the cache misses, and the corresponding cache latency, is to implement prefetchers that have information about the graph processing algorithm beforehand. These prefetchers would then ensure that the appropriate cache blocks are ready in the caches.\\ 
An alternative method would be to improve cache locality by means of graph reordering, in which the graph vertices are reassigned indices without changing the actual structure of the graph itself. Previous studies on different reordering schemes have shown that the reordered graph has lower processing time as compared to the original graph. The reordering scheme itself would be based on a particular feature of the graph that can be exploited, leading to the introduction of a reordering time overhead in the end-to-end processing. Graph reordering involves two tradeoffs that need to be balanced.\\
(1)The more information that we wish to derive from the graph for reordering, the higher would be the reordering time, but at the same time, lower would be the processing time post-reordering. State of the art reordering algorithms take into account information such as the degree of the vertices, the nature and the density of connectivity between the vertices, etc. Thus the reordered graph, although having high execution speedups, also encounters a big overhead in reordering time. Various lightweight reordering algorithms have hence been studied that are simply based on exploiting the power-law distribution in graph- the number of vertices that have high connectivity, and are hence likely to be processed many times, comprise of a small fraction of all the vertices. These hot vertices are sparsely distributed across the entire graph. The probability that a cache block would have multiple hot vertices would be low. However, any memory block having a high number of these hot vertices would be more likely to be preserved in the cache, since they are more likely to be reused due to their high temporal locality. Light-weight reordering algorithms ensure that highly connected vertices, henceforth referred to as hot vertices, are reassigned indices closer to each other.\\
(2) Even though this increases the temporal locality in the graph, it severely affects the spatial locality in the graph in certain cases. Real-world graphs are large in size and have connections governed by natural interaction patterns and thus, carry some inherent connectivity information and display community structures. The memory access patterns are greatly governed by these neighbourhood laws. A simple reordering of the graph vertices on the basis of their degree would threaten to disrupt the spatial locality of neighbourhoods in the original graph structure. For graph processing algorithms that are neighbourhood-based, like BFS, decreasing the spatial locality can therefore lead to worse, if not the same, post-reordering execution time.\\
Taking into consideration the two trade-offs, we introduce our locality-based reordering scheme that seeks to benefit from the natural community structures present in the graph as well as increase the temporal locality by placing the hot vertices together. LOrder defines various disjoint but complete localities in the entire graph structure and then carries out coarse-grain reordering of the vertices in each of the localities- the hot vertices are preferentially numbered before the cold vertices. Within the hot/cold block, the vertex order is maintained in the sequence of BFS traversal during definition of the locality. This method of reordering hot and cold vertices within the locality itself preserves the structure in the graph as well as benefits from the low cache misses by keeping the hot vertices together. In the first version, we define the seed vertex for each community as the unvisited vertex in the original graph with the lowest index. In the second version of the implementation, we define communities as the connected components in the graph that confer to the definition of a ground-truth community in the graph- although the reordering time is higher, we get lower execution graph processing times post-reordering. 
This paper follows this structure
\begin{itemize}
    \item We first study the nuances of graph reordering and criticise some of the existing reordering schemes, including the state-of-art GOrder
    \item We propose LOrder, a new light-weight reordering scheme that carries out localised coarse-grained reordering to improve the spatio-temporal locality in the graph
    \item We compare the results obtained from LOrder with other reordering schemes to show that it has better performance game for many of the cases (beats DBG 77\% of the time) and has a maximum speed up of $7 \times$. We also remark on the role played by the graph diameter on performance
\end{itemize}

\clearpage
\chapter{Background}
\label{sec:Literature_Survey}
\section{Structure of Natural Graphs}
Real-world graphs are large in size and have connections governed by natural interaction patterns and thus, carry some inherent connectivity information and display community structures. The memory access patterns are greatly governed by these neighbourhood laws. These natural graphs can be used to represent many of our network applications like social networks, computer networks, maps, etc. Two prevalent properties of natural graphs that we borrow from-\\
\textbf{Power-law distribution}: Only a small fraction of the vertices have high degree of connectivity. This skewed distribution of degree is an important property with respect to cache reuse since a few of these hot vertices have a high chane of reuse, since they are neighbours of a high number of vertices. A memory block with many hot vertices would likely be queried large number of time, and if it were maintained in the cache, the cache miss rate could be reduced. Table 1 quantifies this skew for the datasets used in this paper. The table contains the percentage of hot vertices in the entire graphs along with the number of edges that come in or out of these vertices. Here, hotness is defined as a vertex having a degree higher than the threshold of average hotness of the vertices for the graph.\\
\textbf{Community structures}: Graphs based on natural connection patterns display high inter-connections in different regions. These clusters are called communities in a graph. For example, in a social network, members sharing similar likes and preferences are bound to converge into groups. The corresponding graph would represent these groups as a closely-knit community of vertices corresponding to these members. In geographical graphs used for navigation purposes, smaller towns would be positioned around important towns or cities.

\section{Graph Representation}
For the purposes of representation, graphs are modelled by a set of vertices \texttt{(V)} and the set of their corresponding directed edges \texttt{(E)}, \texttt{G = (V, E)}. Each of these vertices would have some properties associated with it, which form the basis of graph analysis.\\
In a natural graph, the edges are not uniformly distributed across the vertices but are in fact sparse due to the underlying community structures in the graph. Thus, in an adjacency matrix form of representing graphs, most of the elements would be null, leading to a wastage of expensive shared-memory space.

\begin{figure}
\centering
\begin{subfigure}{.5\textwidth}
  \centering
  \includegraphics[width=.5\linewidth]{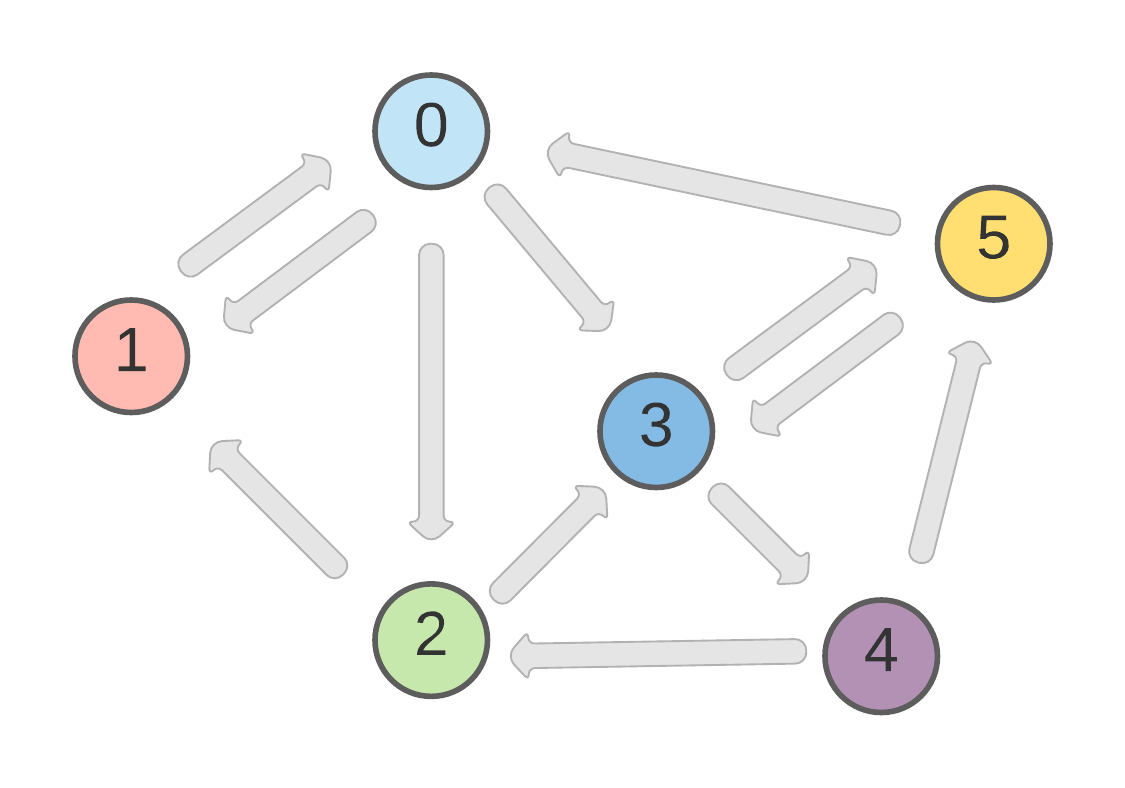}
  \caption{Graph structure}
  \label{fig:sub1}
\end{subfigure}%
\begin{subfigure}{.5\textwidth}
  \centering
  \includegraphics[width=1\linewidth]{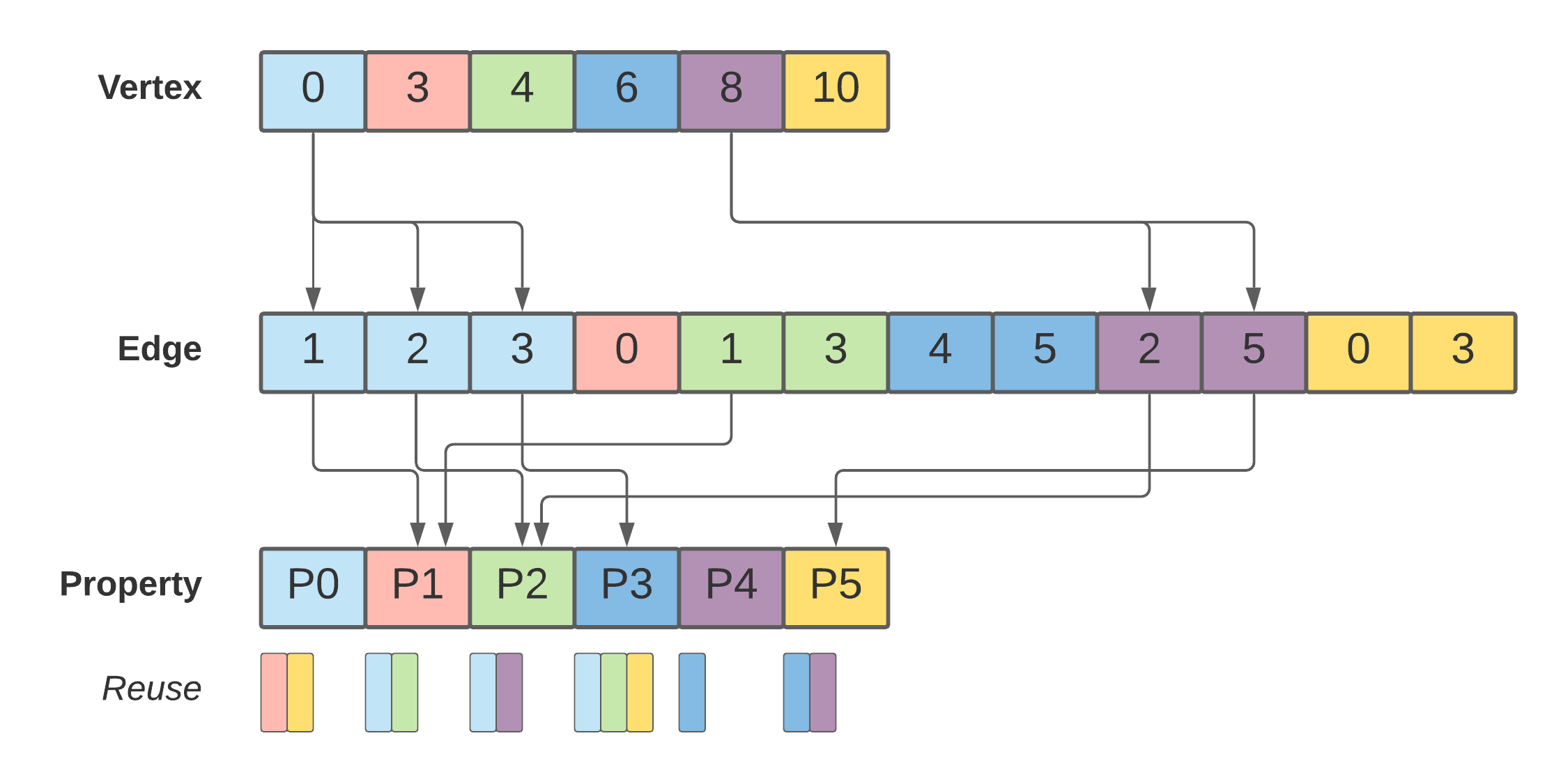}
  \caption{Corresponding CSR format}
  \label{fig:sub2}
\end{subfigure}
\caption{Example of Compressed Sparse Row (CSR) representation of graphs}
\label{fig:test}
\end{figure}

An alternative is found in the Compressed Sparse Row (CSR) format for representing graphs. The CSR format uses a vertex and an edge array to encode the graph, with an additional array to store the properties corresponding to each of the vertices. The edge array could represent an in-edge or an out-edge (in Figure 2.2.1, the out-edges are considered) depending on the type of computations being carried out. In pull-type computations, the parent vertex collects the property values from its in-neighbours. Hence, the CSR format should maintain the in-edges for pull-type computations. In push-type computations, the parent vertex pushes its property values to its out-neighbours for which, the CSR format should maintain the out-edges in the edge array.\\
How is the CSR format queried? We take a pull-type algorithm as an example. For every vertex, the vertex array element corresponding to the vertex id contains the index of the first in-neighbour of the vertex in the edge array. Thus, each element in the edge array would be the source vertex of that edge. The number of edges (the degree) for any vertex can be easily determined by the difference of consecutive elements in the vertex array. In the figure 2.2.1, the \textit{Resuse} demonstrates the number of queries to the property element for each graph vertex for one graph traversal. The number is equivalent to the number of in-edges for the corresponding vertices.

\section{Graph Traversal and Memory Use}
For processing a vertex, the graph processing algorithm would query all its neighbours from the edge array. The property arrays corresponding to each of these neighbours would be queries to process the parent vertex. In the traversal of the entire graph, thus, elements in the vertex and the edge arrays would be queried only once. However, elements in the property arrays would be queried multiple times, depending on the degree of the vertex. Note that this degree would depend on the type of computations: out-degree for pull-based computations and vice versa. The number of reuses of each of the elements can be understood from the figure given. Hence, out of the three arrays, the property arrays show temporal locality. In order to improve the cache utilization from these high-reuse vertices, we try to keep the hotter vertices together. Graph reordering provides the means to optimize the CPU access patterns without affecting the graph structure or the algorithm being executed. The differences in the cache utilizations can be easier understood by observing the below two conditions:

\clearpage
\chapter{Motivation}
\label{sec:Literature_Survey}
 
\section{Statement}
The problem statement of graph reordering as follows:\\
\textbf{Problem statement}: Given that we have a graph modelled as \texttt{G = (V, E)}, we need a permutation function phi() that assigns an unique vertex ID in \texttt{(1, 2, 3…, n)} to each vertex of the original graph, where \texttt{n = mod (V)}, the size of Vertex data structure. The optimal permutation function $\phi()$ should be such that it minimises the total cache misses during execution of a typical graph processing algorithm\\\\
The problem statement can be divided into three objective points for designing, testing and comparing any graph reordering scheme:
\begin{enumerate}
    \item \textbf{Low reordering overhead}- Although our objective is to minimise the total execution time for the graph processing algorithm, we need to also consider the end-to-end time including the time to reorder the entire graph. However, we need not reorder the graph every time we need to carry out any graph analysis, since adding a few vertices to the graph do not cause significant changes in the structure or degrees in the graph. Hence, the reordering time itself can be amortized over multiple instances. The number of traversals that would be required to amortize the reordering time depends on the total reordering time, which thus needs to be curtailed
    \item \textbf{High cache utilization}- keeping as many hot vertices together as possible would lead to better temporal locality of vertices and would thus help in better cache utilization for memory blocks having these hot vertices
    \item \textbf{Preserving the structure}- Many graph analysis algorithms involve high spatio-temporal locality, i.e, they tend to process those vertices first that are in the neighbourhood of the current vertices. Hence, the underlying structures present in the graphs are crucial for determining the data flow in the data flow. Any reordering scheme should preserve these structures to prevent adverse impacts on the performance achieved from improving the temporal locality alone
\end{enumerate}
We now criticise some reordering schemes on the above mentioned objective points.

\section{Graph Reordering (GOrder)}
In GOrder, the authors observe that the data access patterns are greatly influenced by the neighbourhood relations in the graph. In addition to the dependency of the neighbouring vertices on the current vertex, it also observes that the relations between the vertices, called sibling relationships, are of consequence. \\
To formalise the graph reordering problem statement, GOrder takes help of a score function to measure closeness between any two vertices:\\
\begin{equation}
    \mathcal{S}(u,v) = \mathcal{S}_s(u,v) + \mathcal{S}_n(u,v)
\end{equation}
Where,\\ $\mathcal{S}_s$: Number of common in-neighbours of $u$ and $v$\\
$\mathcal{S}_n$: number of times $u$ and $v$ are direct neighbours\\
Thus, two vertices that are densely connected with each other would have a greater score function. If by some means, say a permutation function \texttt{$\phi(.)$}, we ensure that such pairs of vertices are kept close to each other in the cache, the number of cache hits would increase. Mathematically, Gorder defines an accumulated locality score $\mathcal{F}(.)$ over a window of size $\omega$, that should be maximised:
\begin{equation}
    \mathcal{F}(\phi) =  \sum_{0 < \phi(v) - \phi(u) \le \omega}^{} \mathcal{S}(u,v) 
\end{equation}
Maximising of $\mathcal{F}$ is proved to be NP-hard by Gorder, which proposes a greedy algorithm as a solution. Although this greedy algorithm has a high reordering overhead, the execution time achieved after the reordering is considerably lesser. A closer analysis of the results from Gorder suggests that it reorders the hot vertices first as well as keeps the neighbouring vertices together. The combination of these three results make GOrder a very good basis for comparison.

\section{Structure-preserved Reordering (S-Order)}
It proposes the concept of a hypernode, an aggregate of adjacent unvisited cold vertices beginning from a seed vertex into a virtual vertex. The adjacency is defined to be the vertices that are within $\kappa$-hops from the seed vertices. The neighbours of the vertices present in the hypernode are then split into two groups based on their hotness. The reordering begins with the vertices in the hypernode, then the hot neighbours of the hypernode and finally the cold neighbours. 

\section{Neighbourhood Reordering (N-Order)}
Norder takes a very different approach to balancing the tradeoffs. It first creates a list of vertices in descending order of their hotness. The next iteration, it carries our reordering in BFS-fashion serially taking elements from the previously arranged list as seed vertices for the BFS. The new reordering is in the order of traversal. Since the entire graph would be traversed twice during NOrdering, the reordering time is expected to be proportionally high as well, as discussed later during the results comparison.

\section{Degree-Based Grouping (DBG)}
DBG is a skew-aware coarse-reordering technique that rearranges the graph such that each cache block has vertices having similar degrees. As opposed to Sort, which just reorders the entire graph with new vertex ids being assigned in descending order of their hotness, DBG bins the vertices in different partitions, in the original vertex numbering order, i.e in each of the partitions, DBG maintains the original relative order of the vertices. But that is the extent to which it ensures the integrity of the original structure of the graph. The partitions are defined with degree ranges that follow the power-law distribution. Since this technique involves only ‘binning’ and not a full-blown sorting of the vertices, the reordering time incurred by DBG is found to be the least for most of the cases. However, since it conforms to the power-law distribution directly, the results from DBG reordering are quite aligned with our objective points. 
\\\\Some key observations after studying the above schemes:
\begin{enumerate}
    \item Since one of the objectives is to maintain the neighbourhood structures of the original graph, many of the reordering algorithms thus make use of BFS-type pattern to traverse the entire graph for indexing
    \item Although GOrder shows promising results for execution time, the reordering time is quite high
    \item By forming hypernodes, SOrder tends to prefer preserving neighbourhood structure over improving the temporal locality
    \item Norder, on the other hand, trades off the temporal locality over preserving the community structure, while also encountering significant reordering time due to the double traversal
    \item Design of DBG limits its structure-preservation to just maintaining the relative order of the reordered vertices. The serial vertex IDs of the original graph may not be the best representative of the underlying cross connections in the graph
\end{enumerate}

\clearpage
\chapter{Locality-based Reordering}

To acknowledge the inherent tendency in the vertices to be a part of communities, we propose Lorder that aims to define small localities in the graph and carry out reordering keeping these localities as the basis. By reordering localities in their entirety and then a coarse-grain sorting of vertices within the localities, Lorder promises to well balance the trade-off between improving spatial and temporal locality. \\
Lorder defines a locality as all the vertices that fall within a $\kappa$-hop distance from a seed vertex, traversed in BFS fashion. The localities are disjoint and thus, their formation and structure greatly depend on the first seed vertex that is chosen. We also assign hotness to each of the localities, henceforth called locality hotness, as the number of hot vertices part of the locality. Hot vertices are themselves defined to be those vertices having degrees more than the average degree of the entire graph. The reordering process can be thus divided into two different parts.\\\\
\textbf{Locality formation}
At the primitive stage, we use the zeroth vertex of the original graph as the seed vertex, which also forms a tag for the locality. Hence, all the vertices within $\kappa$-hops from the zeroth vertex become a part of its locality. As vertices are added to the locality, a count of the hot vertices in it is maintained and the vertices themselves are marked with the locality ID (the reason discussed later).\\
After the locality is completely formed, the tag of the locality along with the corresponding hotness is maintained in a list LocalityID. We serially take up remaining unvisited vertices to be the seed vertex and repeat the above mentioned steps. Once all the vertices are visited and are thus part of a unique locality, we sort the LocalityID list in descending order of the locality hotness. This concludes the locality formation part of the algorithm, summarised as Algorithm 1.\\
\textbf{Assigning new indices}
Once we have the sorted LocalityID list, we begin with the actual assignment of new indices. Taking each seed vertex serially, we begin with traversing the vertices that are within $\kappa$-hops from it and are a part of its locality. However, as earlier mentioned, the distribution of vertices into the localities depends on the seed vertex we start with. Hence, to ensure consistency between localities defined in Algorithm 1 and the vertices being visited this time, we associate each vertex with the tag of the locality it belongs with.\\
For each locality, we maintain a hot and a cold vertices list. As we traverse the locality from the seed vertex, we serially add vertices in their respective list. Lorder preferentially assigns consecutive indices to the hot vertices, before the cold vertices. The process is summarised in Algorithm 2.
\begin{algorithm}
\DontPrintSemicolon
  
  \KwInput{Graph \texttt{G = (V, E)}}
  \KwOutput{Original graph \texttt{G = (V, E)} and the list \texttt{LocalityID} with locality seed vertex IDs and corresponding community hotness}
  %\KwData{Testing set $x$}
  %$\sum_{i=1}^{\infty} := 0$ \tcp*{this is a comment}
  
  \For{$v \in$ V}{
    \If {!Assigned}{
        $seed$ = $v$, $list$ = $\{$v$\}$; $hops\_left = \kappa$; \\
        \While{$hops\_left$ != 0 }{
        $hops\_left$- -;\\
        
            \For{ i $\in Neigh_{out}(v)$ $\&$ !visited(i)}{
            localityID(i) = v\\
            visited(i) = true\\
            list.append(i)
            }

        }

    }

  }

\caption{Formation of localities}
\end{algorithm}
\begin{algorithm}
\DontPrintSemicolon
  
  \KwInput{Graph \texttt{G = (V, E)}}
  \KwOutput{Graph \texttt{G = (V, E)}, list \texttt{LocalityID}}
  seed = 0, visited() reset, Assigned() reset, index =0;\\
  \For{$v \in$ V}{
    \If {!Assigned}{
        $seed$ = $v$, $list$ = $\{$v$\}$; $hops\_left$ = $\kappa$; \\
        new$\_$id[seed] = index++;\\
        \While{$hops\_left$ != 0 }{
        $hops\_left$- -;
            \For{ i $\in Neigh_{out}(v)$ $\&$ !visited(i)}{
                \If{localityID(i) == v}{
                new$\_$id[i] == index++;\\
                visited[i] = true;
                }
            }
        }
    }
  }
\caption{Assigning new ids to vertices}
\end{algorithm}

\begin{figure}
\centering
\begin{subfigure}{.5\textwidth}
  \centering
  \includegraphics[width=.9\linewidth]{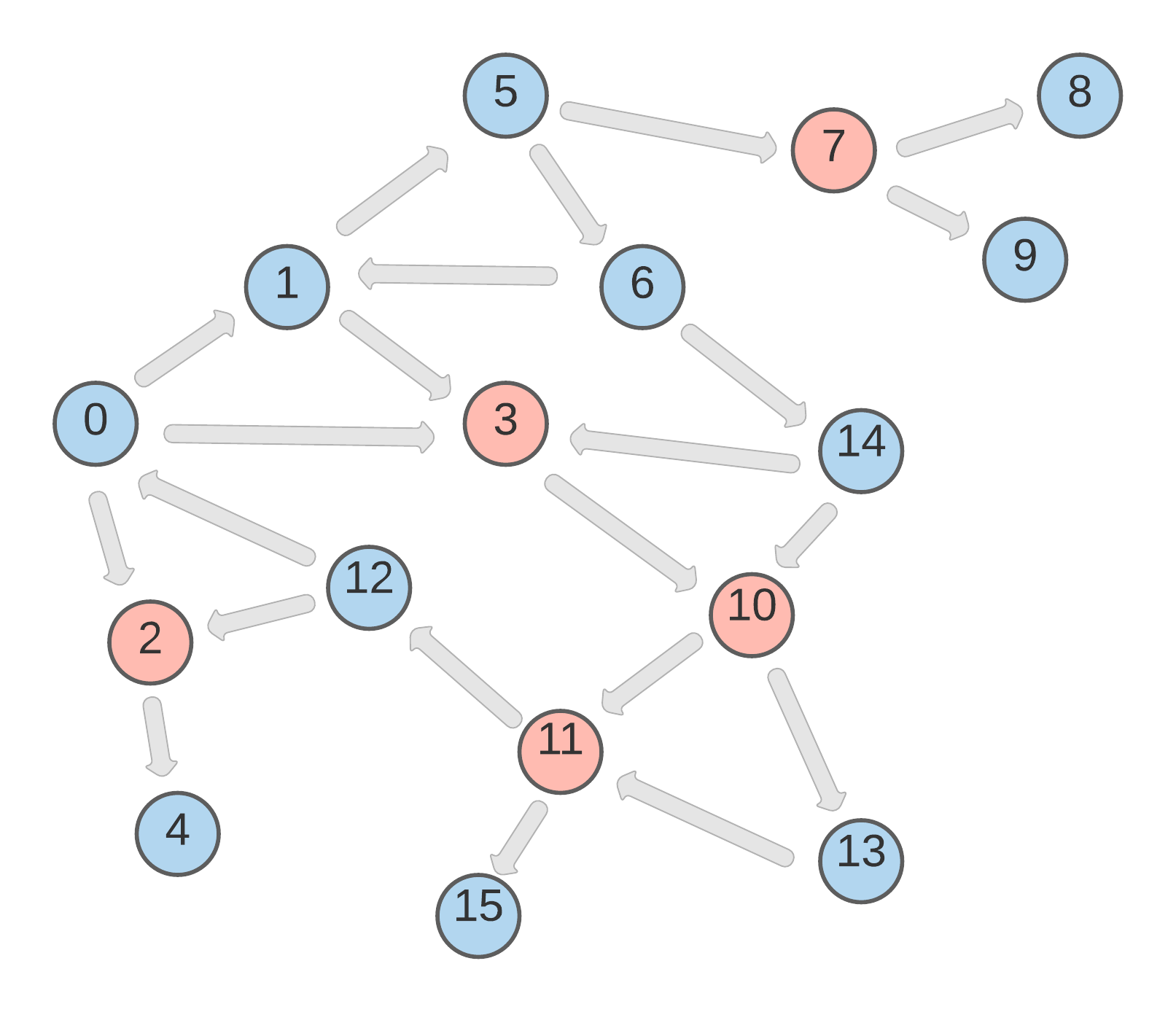}
  \caption{Original graph structure}
  \label{fig:sub1}
\end{subfigure}%
\begin{subfigure}{.5\textwidth}
  \centering
  \includegraphics[width=1\linewidth]{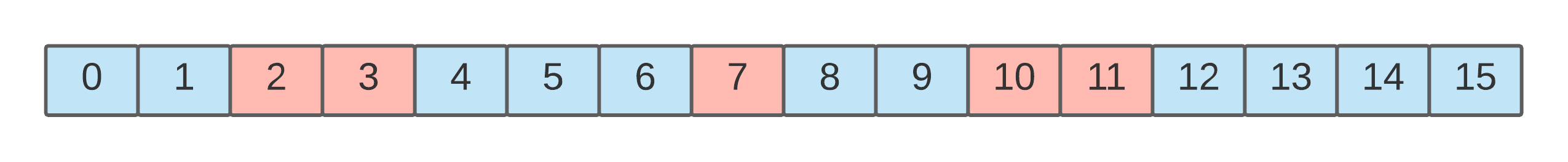}
  \caption{Property array}
  \label{fig:sub2}
\end{subfigure}
\caption{(b) shows the arrangement of the property array in the memory for graph in (a)}
\label{fig:test}
\end{figure}

\begin{figure}
\centering
\begin{subfigure}{.5\textwidth}
  \centering
  \includegraphics[width=.9\linewidth]{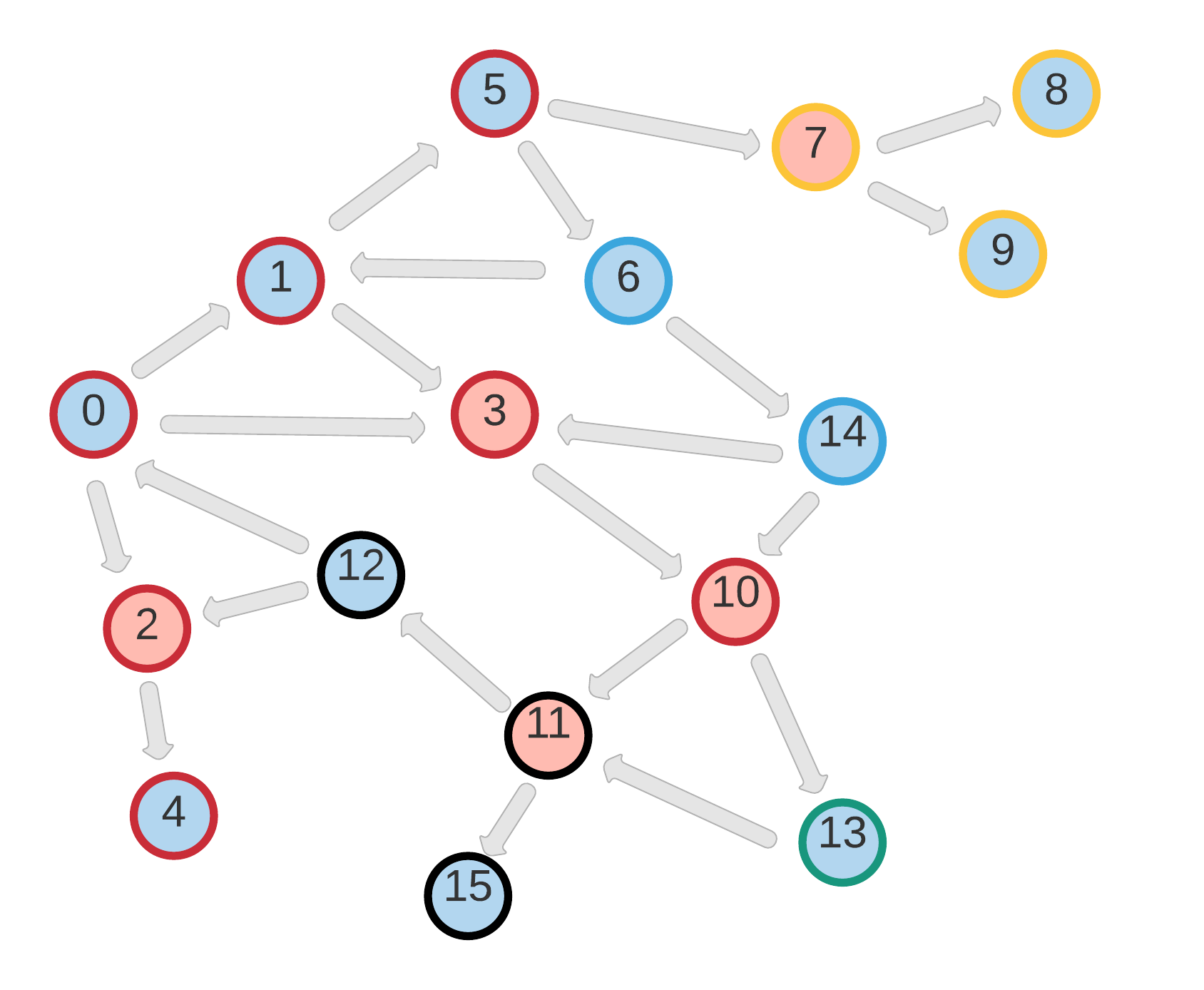}
  \caption{Localities}
  \label{fig:sub1}
\end{subfigure}%
\begin{subfigure}{.5\textwidth}
  \centering
  \includegraphics[width=0.75\linewidth]{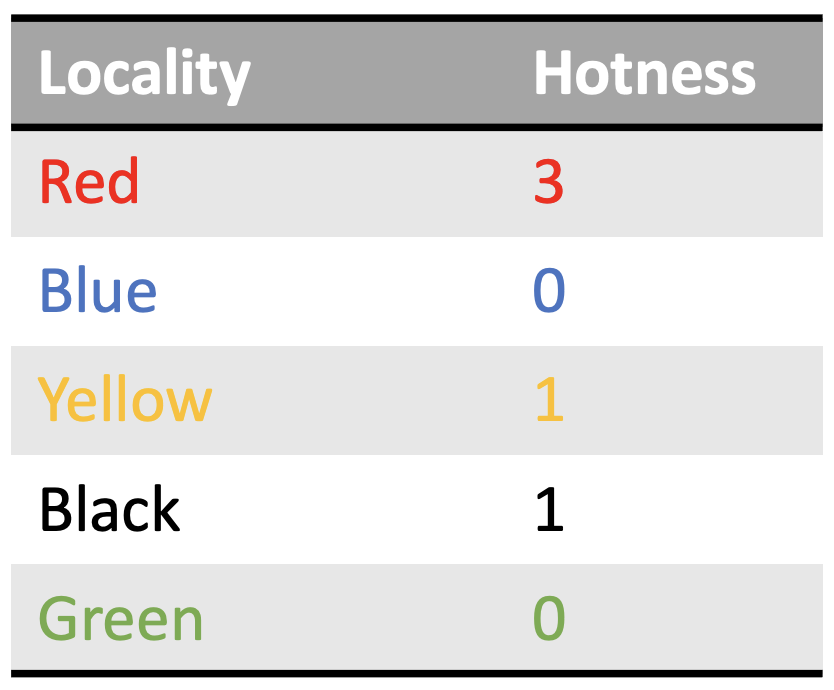}
  \caption{Locality-hotness}
  \label{fig:sub2}
\end{subfigure}
\caption{(b) shows the arrangement of the property array in the memory for graph in (a)}
\label{fig:test}
\end{figure}

\begin{figure}
\centering
\begin{subfigure}{.5\textwidth}
  \centering
  \includegraphics[width=.9\linewidth]{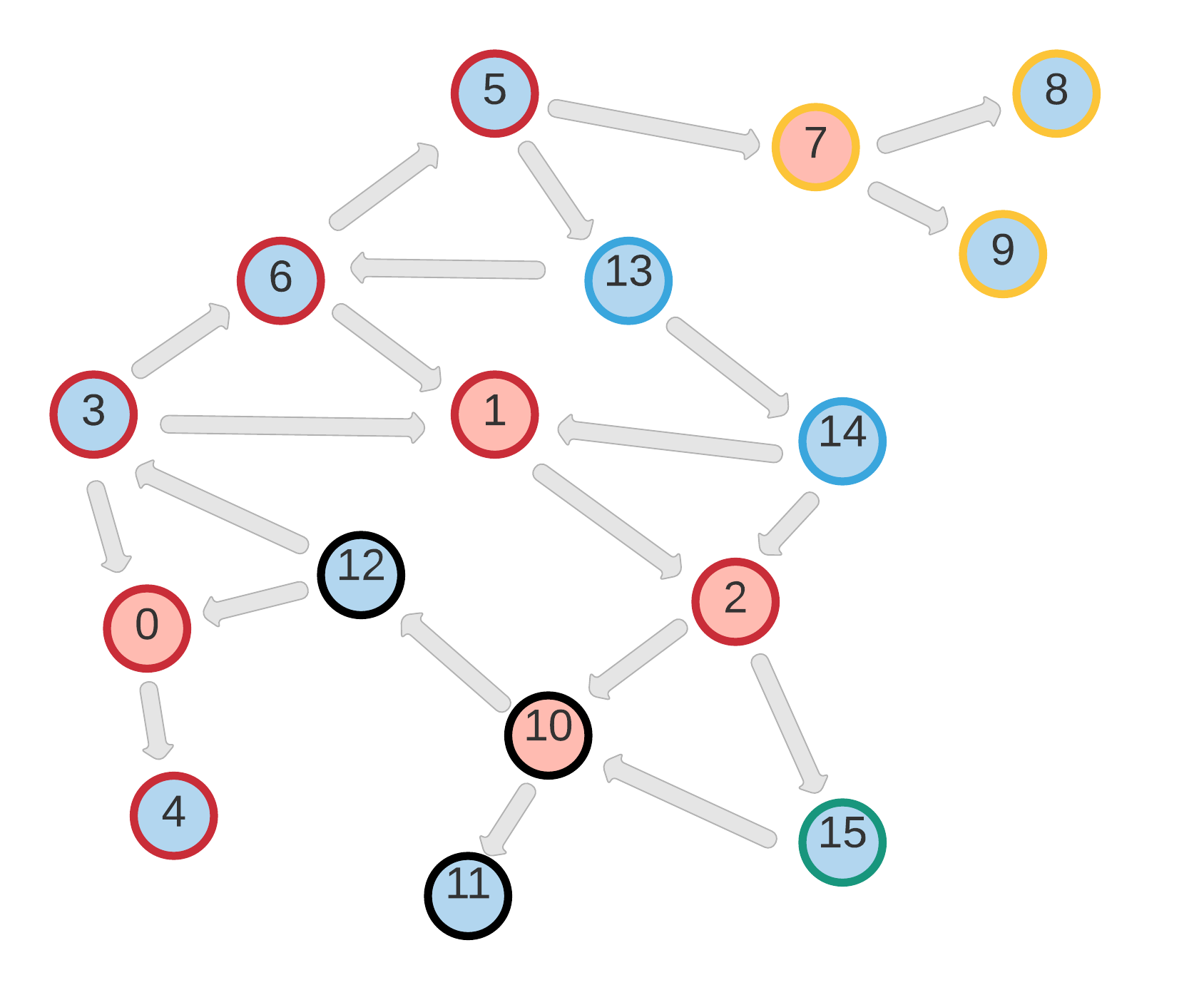}
  \caption{Reordered graph}
  \label{fig:sub1}
\end{subfigure}%
\begin{subfigure}{.5\textwidth}
  \centering
  \includegraphics[width=1\linewidth]{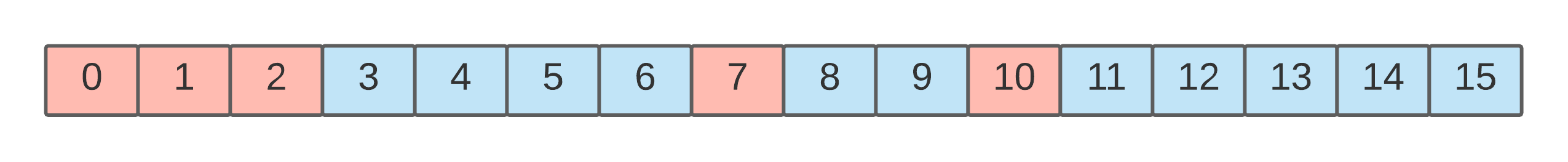}
  \caption{Property array}
  \label{fig:sub2}
\end{subfigure}
\caption{(b) shows the arrangement of the property array in the memory for graph in (a)}
\label{fig:test}
\end{figure}

\clearpage
\chapter{Performance Evaluation}
\section{Experimental Setup}
We evaluate effectiveness of Lorder by running six different graph processing algorithms on graph reordered by four graph reordering schemes. The evaluations are run on a machine with following configuration: (add compiler version here too) \\
\textbf{Algorithms}: We make use of the GAPS Benchmark suites for carrying out the graph processing. It has C++ implementations of the following algorithms:
\begin{itemize}
    \item \textit{Breadth-First Search}: A fundamental graph traversal technique which starts from a single source and traverses all the vertices of the current level before moving on to the next one
    \item \textit{Page Rank}: Iteratively computes the rank of vertices using their connectivity propoerties
    \item \textit{Betweenness centrality}: Using BFS traversal to determine the paths originating from a aprticular source through a particular vertex, it determines a central vertex
    \item \textit{Single Source Shortest Path}: Using the Bellman Ford algorithm, it finds shortest path to all the connected vertices from a particular vertex
    \item \textit{Connected Components}: It assigns unique labels to all the connected components and the vertices associated with each of these components
    \item \textit{CC\_SV}: Connected Components implemented with the Shiloach-Vishkin algorithm
\end{itemize}

\begin{table}
\centering
\begin{tabular}{llll} 
\hline
\rowcolor[rgb]{0.502,0.502,0.502} \multicolumn{1}{c}{\textbf{\textcolor{white}{Dataset}}} & \multicolumn{1}{c}{\textbf{\textcolor{white}{Type}}} & \multicolumn{1}{c}{\textbf{\textcolor{white}{~\# of vertices~}}} & \multicolumn{1}{c}{\textbf{\textcolor{white}{~\# of nodes~}}}  \\ 
\hline
\rowcolor[rgb]{0.753,0.753,0.753} lj                                                      & Social Network                                       & 4,847,571                                                        & 68,993,773                                                     \\
orkut                                                                                     & Community                                            & 3,072,441                                                        & 117,185,083                                                    \\
\rowcolor[rgb]{0.753,0.753,0.753} pld arc                                                 & Hyperlink                                            &                                                                  &                                                                \\
kron23                                                                                    & Kronecker                                            &                                                                  &                                                                \\
\rowcolor[rgb]{0.753,0.753,0.753} youtube                                                 & Community                                            & 1,134,890                                                        & 2,987,624                                                      \\
pokec                                                                                     & Social Network                                       & 1,632,803                                                        & 30,622,564                                                    
\end{tabular}
\caption{Dataset details}
\label{table:graphsize}
\end{table}

\textbf{Graph datasets}: A total of six graph datasets, summaried in Table XYZ. Apart from the Kronecker graph by Graph500 benchmark, all the other graphs are real-world graphs. The average degrees of these graphs are also mentioned, which forms the threshold for defining hotness for each vertex. We also report the diameter of the graphs, the longest shortest connected path in the graph, for those belonging to SNAP Stanford dataset repository. The diameter plays an integral part in the analysis of Lorder, as summarised in the results.\\
\textbf{Evaluation Method}
For the various reordering schemes, the parameters are set as per the inferences by their authors. For example, we set $\lambda = 50$ and $\kappa = 2$ for SOrder. For LOrder, we set $\lambda = $ the average degree of the graph. We also have the $\kappa$ parameter in Lorder, which varies from graph to graaph. the value is chosen such that the final execution time of the algorithms is least for the set $\kappa$ value. The comparison across the reordering schemes are based on the execution speedups, the cache statistics, reordering time as well as interesting observation about the $\kappa$ values. We report execution times averaged over 16 iterations to incorporate time required for cache warm-up.

\section{Results}
\begin{figure}
    \centering
    \includegraphics[scale =0.5]{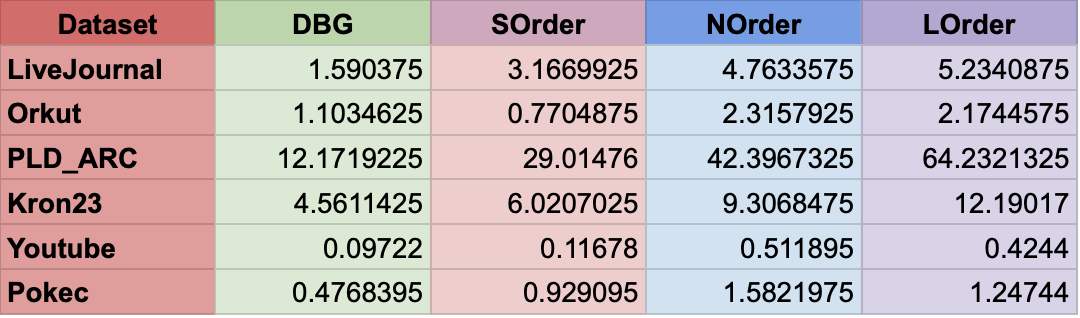}
    \caption{Time taken (in seconds) to reorder graphs with each of the four reordering schemes}
    \label{fig:Reordering}
\end{figure}

\begin{figure}
\begin{subfigure}{\textwidth}
    \centering
    \includegraphics[scale=0.4]{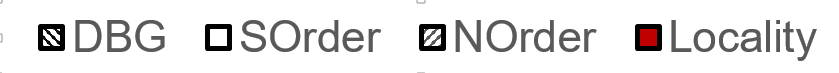}
\end{subfigure}

\begin{subfigure}{.5\textwidth}
  \centering
  \includegraphics[width=.8\linewidth]{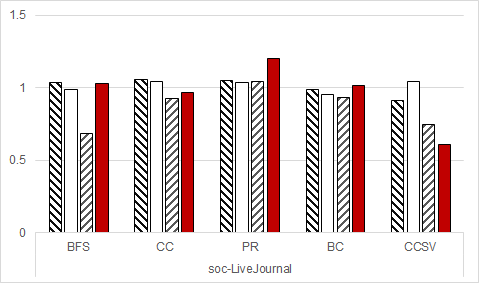}
  \caption{soc-LiveJournal1}
  \label{fig:sfig1}
\end{subfigure}%
\begin{subfigure}{.5\textwidth}
  \centering
  \includegraphics[width=.8\linewidth]{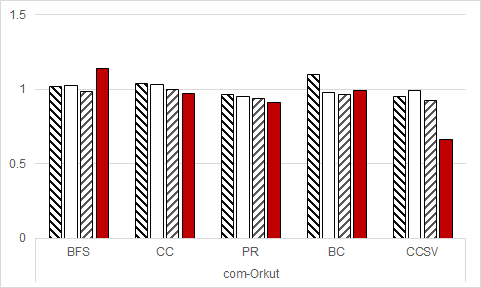}
  \caption{com-Orkut}
  \label{fig:sfig2}
\end{subfigure}
\begin{subfigure}{.5\textwidth}
  \centering
  \includegraphics[width=.8\linewidth]{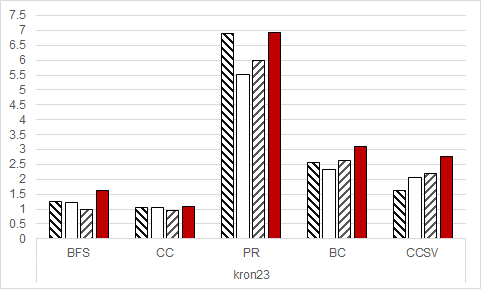}
  \caption{kron23}
  \label{fig:sfig3}
\end{subfigure}%
\begin{subfigure}{.5\textwidth}
  \centering
  \includegraphics[width=.8\linewidth]{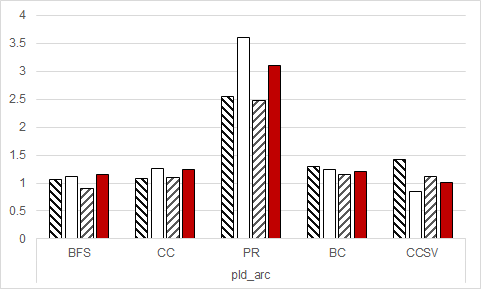}
  \caption{pld\_arc}
  \label{fig:sfig4}
\end{subfigure}
\begin{subfigure}{.5\textwidth}
  \centering
  \includegraphics[width=.8\linewidth]{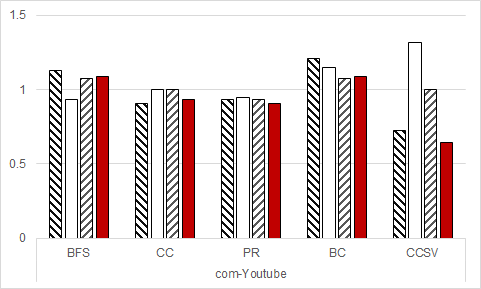}
  \caption{com-Youtube}
  \label{fig:sfig5}
\end{subfigure}%
\begin{subfigure}{.5\textwidth}
  \centering
  \includegraphics[width=.8\linewidth]{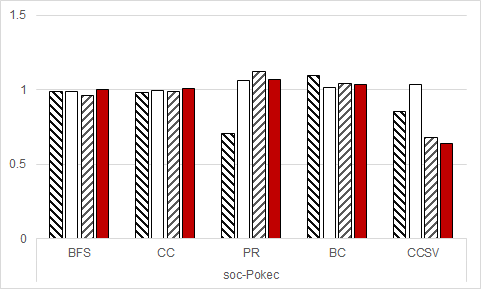}
  \caption{soc-Pokec}
  \label{fig:sfig2}
\end{subfigure}
\caption{Speed ups comparison contributed by different reordering algorithms for various datasets. It can be seen that a basic implementation of Lorder without any embellishments performs better in many cases as discussed in the results}
\label{fig:fig}
\end{figure}

\begin{figure}
    \centering
    \includegraphics{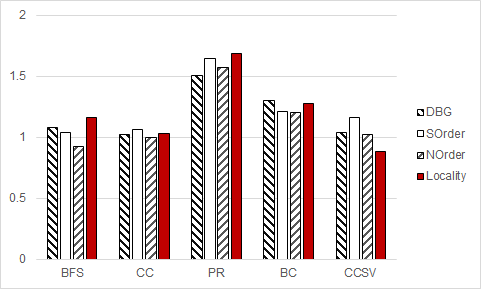}
    \caption{Geometric means of graph processing speed-ups for various kernels across different datasets}
    \label{fig:GMean}
\end{figure}

\textbf{The value of $\kappa$}. The number of hops from the seed vertex, parameter $\kappa$, determines the locality for any seed vertex. The value is set such that the post-reordering execution time is minimum. The results are as shown in the Table \ref{table:dataset}. The value of $\kappa$ is observed to be half of the diameter of the graph, as mentioned on the SNAP dataset repository. Hence, $\kappa$ is also referred to as the \textbf{\textit{radius}} in this paper. For the remaining datasets, the diamater can be shown to be twice of the radius.
\begin{table}
\centering
\begin{tabular}{llll} 
\hline
\rowcolor[rgb]{0.502,0.502,0.502} \multicolumn{1}{c}{\textbf{\textcolor{white}{Dataset}}} & \multicolumn{1}{c}{\textbf{\textcolor{white}{Type}}} & \multicolumn{1}{c}{\textcolor{white}{\textbf{~Diameter of the graph}}} & \multicolumn{1}{c}{\textbf{\textcolor{white}{~Value of $\kappa$ ~}}}  \\ 
\hline
\rowcolor[rgb]{0.753,0.753,0.753} lj                                                      & Social Network                                       & 16                                                                     & 8                                                                                     \\
orkut                                                                                     & Community                                            & 9                                                                      & 4                                                                                     \\
\rowcolor[rgb]{0.753,0.753,0.753} youtube                                                 & Community                                            & 20                                                                     & 10                                                                                    \\
pokec                                                                                     & Social Network                                       & 11                                                                     & 5                                                                                    
\end{tabular}
\caption{The value of $\kappa$ is set to achieve the lowest execution time}
\label{table:dataset}
\end{table}
\\\textbf{Comparing processing speed-up}
The proposed reordering scheme is implemented on the GAPS Benchmark and the execution times post-reordering are compared to the execution time without reordering. The speed-ups are plotted in Figure \ref{fig:fig} for six different datasets and for five graph processing kernels. Since DBG provides some of the best processing speed-ups, we compare our results to see that \textbf{Lorder beats DBG for 77\% of the total comparisons} (i.e, 23 times out of 30). Likewise, it beats SOrder 60\% of the time (i.e, 18 times out of 30). To better analyse the overall behaviour of the reordering schemes, we plot their geometric means for different graph processing kernels in Figure \ref{fig:GMean}. We can see that Lorder outright beats all the other reordering schemes for BFS, CC and PR applications while it is marginally slower that DBG in BC. However, Lorder consistently performs badly for the CC (Shiloach-Vishkin) graph kernel. This can be attributed to the uncertainty in which vertices would be a part of the processed at any given time.\\
It is also observed that for apart for a few cases, reordering graphs lead to significant processing speed-ups. For applications involving multiple iterations or traversals, these speed-ups can greatly reduce the total processing time. However, we also need to account for the reordering times.\\
\textbf{Comparing reordering times}. Out of the four reordering schemes being discussed, DBG and SOrder require a single traversal of the graph while NOrder and LOrder require two graph traversals. So it is expected of the latter two to have almost double the reordering time of the former, as collaborated through Table \ref{fig:Reordering}. The size of the graph also determines the time taken to reorder it. The values in the above table follow the graph size trends in Table \ref{table:graphsize}.\\
However, the actual reordering time also depends on factors like the number of hot vertices being processed initially, etc. The cost of reordering is not incurred every iteration of processing. A reordered graph is stored in the memory and thus, can be used multiple times. Also, the addition and removal of vertices to the reordered graph with significantly high number of vertices does not affect the processing time much, as shown by a previous study. Hence, the reordering time is amortized over multiple iterations.

%\end{document}
%\input{./Chapters/Chapter_6}
%\input{./Chapters/Chapter_7}
%\input{./Chapters/Chapter_8}

%%%%%%%%%%%%%%%%%%%%%%%%%%%%%%%%%%%%%%%%%%%%%%%%%%%%%%%%%%%%%
%APPENDICES
%%%%%%%%%%%%%%%%%%%%%%%%%%%%%%%%%%%%%%%%%%%%%%%%%%%%%%%%%%%%%

%%%%%%%%%%%%%%%%%%%%%%%%%%%%%%%%%%%%%%%%%%%%%%%%%%%%%%%%%%%%%
%BIBLIOGRAPHY
%%%%%%%%%%%%%%%%%%%%%%%%%%%%%%%%%%%%%%%%%%%%%%%%%%%%%%%%%%%%%

% \clearpage
%\renewcommand*{\thesection}{}\textbf{}
%\label{Bibliography}
%\include{bib_kp}

\label{References}
%\lhead{\emph{Bibliography}}

%\begin{thebibliography}{15}

%\bibitem{ref_10}
%https://www.researchgate.net/figure/A-simplified-view-of-the-RoCC-interface_fig7_324722826

%All the references will be added soon

%end{thebibliography}

% \bibliographystyle{abbrv}

% \bibliography{bibliography}

\end{document}